\documentstyle[twocolumn,prl,epsfig,aps,floats]{revtex}
\tightenlines

\def\SrCu{Sr$_{14}$Cu$_{24}$O$_{41}$ }

\def\Sr2.5{Sr$_{2.5}$Ca$_{11.5}$Cu$_{24}$O$_{41}$ }

\def\SrCuOCl{Sr$_{2}$CuO$_{2}$Cl$_{2}$}

\def\cm-1{cm$^{-1}$}

\begin{document}
\twocolumn[
\hsize\textwidth\columnwidth\hsize\csname@twocolumnfalse\endcsname
\draft
\title
{Spin dynamics of \SrCu two-leg ladder studied by Raman
spectroscopy
}

\author{
A. Gozar$^{1,2}$, G.~Blumberg$^{1,\dag}$,
B.S. Dennis$^{1}$, B.S. Shastry$^{1}$,
N. Motoyama$^{3}$, H. Eisaki$^{4}$, and S. Uchida$^{3}$}

\address{
$^{1}$Bell Laboratories, Lucent Technologies, Murray Hill, NJ 07974 \\
$^{2}$University of Illinois at Urbana-Champaign, Urbana, IL 
61801-3080 \\
$^{3}$The University of Tokyo, Bunkyo-ku, Tokyo 113, Japan\\
$^{4}$Stanford University, Stanford, CA94305\\
}
\date{April 2, 2001; Accepted for Physical Review Letters}
\maketitle

\begin{abstract}

The two-magnon (2M) excitation at 3000 \cm-1 in \SrCu two-leg
ladder is studied by Raman scattering.
A slight anisotropy of the superexchange coupling $J_{\perp} / J_{||} 
\approx 0.8$ with $J_{||} = 110 \pm 20$~meV is proposed from the 
analysis of the magnetic scattering.
The resonant coupling across the charge transfer gap increases the 2M 
intensity by orders of magnitude.
The anisotropy of Raman scattering is dependent upon the excitation 
energy.
The 2M relaxation is found to be correlated with the temperature
dependent electronic Raman continuum at low frequencies.

\end{abstract}

\pacs{PACS numbers: 74.72.Jt, 78.30.-j, 75.40.Gb, 75.10.Jm}
]
\narrowtext

\emph{Introduction}.--
Spin dynamics of low-dimensional copper
oxide materials with spin S=1/2 copper ions is
attracting much attention because of the critical nature of the ground
state and its possible relevance to the phase diagram of
high temperature superconducting cuprates \cite{Anderson,Fisher}.
Quantum fluctuations in the S=1/2 Heisenberg one-dimensional (1D)
antiferromagnetic (AF)
linear chain $H_{||}=J_{||} \sum_{leg} ({\bf S}_i \cdot {\bf
S}_j - \slantfrac{1}{4})$ are so
strong that the ground state is disordered and
gapless \cite{Faddeev}.
The excitations from the ground state are S=1/2 topological kinks
(solitons) called \emph{spinons}.
Two chain coupling in two-leg ladder structures is described by the
spin Hamiltonian $H = H_{\perp} + H_{||}$,
where $H_{\perp}=J_{\perp} \sum_{rung} {\bf S}_i \cdot {\bf S}_j$
is the interchain coupling.
If the AF coupling across the rung $J_{\perp}$ is much
larger than $J_{||}$, the ground state consists of spin
singlets, one on each rung, with a gap $\Delta \approx
J_{\perp}$ to the lowest triplet spin excitation.
A finite $J_{||}$ will drive the system into a resonating valence
bond (RVB) state \cite{Rice96,Barnes93,Oitmaa,White,Kivelson} with a 
finite
spin gap for any value of the ratio $r =  J_{\perp}/J_{||}$.
The RVB state is a coherent superposition of valence bonds.
For even leg ladders the RVB states are short ranged and can be 
visualized
as predominantly interchain singlets with resonances to include 
intrachain
singlets.
The elementary excitations out of the RVB ground state are $S = 1/2$
quasiparticles that carry no charge.
They are viewed as topological defects that can be created or 
destroyed
only in pairs by breaking a singlet bond.
Gradual introduction of interladder coupling drives the system from
the quantum RVB state to a magnetically
ordered 2D N\'eel state with classical spin-wave excitations 
\cite{Sachdev}.

\SrCu is an experimental realization of a two leg ladder structure.
The planes of Cu$_{2}$O$_{3}$ weakly coupled ladders are stacked
along the $b$ crystallographic axis alternating with 1D CuO$_{2}$
edge-sharing chain sheets \cite{McCarron,Siegrist}.
The legs and rungs of the ladders are along the $c$ and
$a$ crystallographic axes.
This compound is intrinsically doped and the holes are believed to 
reside
mainly in the chain substructure \cite{Osafune97}.
The spin dynamics in the ladders is governed by the ratio $r$ of the
magnetic exchange across and along the legs.
INS and NMR measurements suggest superexchange energies $J_{\perp}
\approx 72 - 80$~meV and $r \approx 0.5$ \cite{Eccleston,Imai} while
Raman measurements derived values close to the isotropic limit for
$J_{\perp}$ and $J_{||}$ \cite{Sugai99}.

In this Letter we study resonant magnetic Raman scattering from \SrCu
which provides information about the spin dynamics from the two-magnon
(2M) peak position and shape \cite{ElliottParkinson}.
The 2M feature, which displays distinct characteristics from the
corresponding excitation in 2D cuprate materials, is analyzed as a
function of temperature, polarization and incoming photon energy.
From the energy of the 2M resonance and the value of the spin gap we 
suggest $r = 0.8 \pm 0.1$ and $J_{||} = 110 \pm 20$~meV.
The temperature dependence of the magnetic scattering is found to be
correlated with the suppression of the low frequency Raman continuum
and the redistribution of states seen in optical absorption,
reflecting that the low-energy spin-dynamics is driven by temperature
dependent high energy interactions.

\emph{Experimental}.--
Single crystals of \SrCu were grown as described in
\cite{Osafune97,Motoyama}.
Raman measurements were performed from the $ac$ surface of the crystal
mounted in a continuous helium flow optical cryostat.
Spectra were taken in a backscattering geometry using  linearly
polarized excitations of a Kr$^{+}$ laser from IR to violet.
An incident laser power less than 3 mW was focused to a 50~$\mu$m
spot onto the sample surface.
The spectra were analyzed by a custom triple grating spectrometer and 
the data were corrected for the spectral response of the spectrometer 
and the detector as well as for the optical properties of the 
material at different wavelengths as described in
\cite{Blumberg94,MotoyamaOptics}.

\emph{Magnetic scattering}.--
In Fig.~1 we present low-temperature magnetic Raman scattering spectra
from \SrCu for (cc) and (aa) polarizations.
The data is taken in the pre-resonant regime (1.84 eV excitation).
The peak at 3000~\cm-1 (375~meV) is assigned to a photon induced spin
exchange or 2M excitation.
The 2M resonance is a sharp asymmetric feature with a tail at higher
frequencies.
In (aa) polarization the peak is about four times weaker than in (cc) 
polarization and is found at the same energy.
A semi-classical counting of broken magnetic bonds within a local
N\'eel
environment as shown in Fig. 2a-b  was proposed to describe the spin
exchange process and determine $J_{||}$ and $J_{\perp}$
\cite{Sugai99}.
For anisotropic coupling along and across the legs a simple Ising
counting
leads to different peak energies in the (aa) and (cc) polarizations
which we did not observe.
The isotropic case estimates $J \approx 200$~meV which is almost
50\% higher than in related cuprate materials.

\begin{figure}[t]
\centerline{
\epsfig{figure=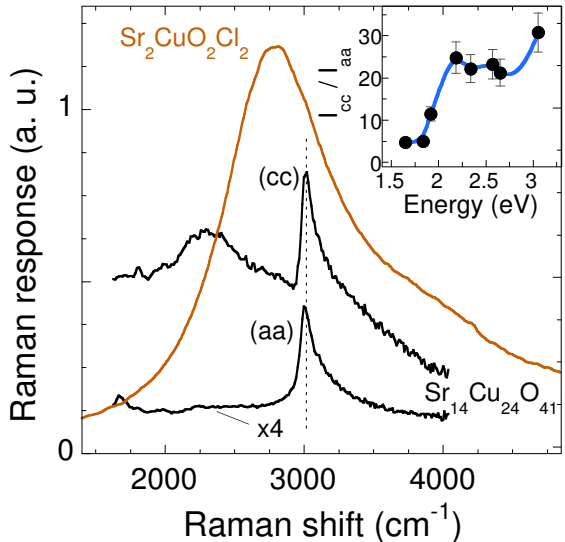,width=75mm}
}
\caption{
Low temperature Raman response function in (cc) and (aa)
polarization for \SrCu with $\omega_{in} = 1.84$~eV excitation.
The broad feature is 2M excitation for 2D AF insulator \SrCuOCl with
$\omega_{in} = 2.73$~eV.
Inset: the relative intensity of the 2M excitation in (cc) $vs.$ (aa) 
polarization as a function of the incoming photon energy (from 
Fig.~4). The solid line is a guide to the eye.
}
\label{Fig.1}
\end{figure}

An RVB description of the ladder ground state has been suggested
\cite{White}.
Fig.~2c-d is a cartoon of the local magnetic excitation within this
approach.
Starting from an "instantaneous" configuration of the ground state, 
(Fig.~2c), during the Raman process two neighboring singlets get 
excited into a higher singlet state of two bound triplets (Fig.~2d).

An effective spin interaction leading to the 2M light scattering was 
first developed by Loudon and Fleury \cite{Fleury}. 
For Mott-Hubbard systems the photon induced spin exchange formalism 
was developed in pre-resonant \cite{Sriram} and resonant 
\cite{Chubukov} regimes. 
In our notation ${\bf e}_{in}$ and ${\bf e}_{out}$ are the 
polarization vectors of the incoming and scattered light, and 
$\theta$ is the the angle between ${\bf e}_{in}$  and the $a$-axis.
For ${\bf e}_{in} || {\bf e}_{out}$, considering nearest neighbor 
exchange and taking into account the anisotropy of the coupling 
constants denoted by $A$ and $B$ \cite{Sugano}, the light scattering 
interaction is:
\begin{equation}
H_{LS} = A \sin^{2}(\theta) \sum_{leg} {\bf S}_i \cdot {\bf S}_j + B 
\cos^{2}(\theta) \sum_{rung} {\bf S}_i \cdot {\bf S}_j
\label{LS}
\end{equation}
Using the relationship between
$H_{LS}$ and the nearest-neighbor Heisenberg ladder Hamiltonian, the 
following angular dependence of the 2M intensity for ${\bf e}_{in} || 
{\bf e}_{out}$ can be found \cite{Freitas2000}:
\begin{equation}
I_{||}(\omega, \theta) = I_{||}(\omega, 0) [\cos^2(\theta) - 
\frac{A}{B}r \sin^2(\theta)]^{2}
\label{Itheta}
\end{equation}
The ratio of the 2M peak intensity in parallel polarization along and 
across the ladder allows the extraction of the exchange anisotropy if 
the $A$ to $B$ ratio is known.
In resonance the $A / B$ value is excitation energy dependent 
and it approaches a constant value in the pre-resonant regime 
\cite{Sriram}.
This can be seen in the inset of Fig.~1 where we show the ratio of 
the 2M intensity for (cc) and (aa) polarizations as a function of 
$\omega_{in}$.

The constraints imposed by the 2M energy at 3000~\cm-1 measured by 
Raman and the spin gap value $\Delta = 32$~meV measured by INS 
\cite{Eccleston} allow us to estimate $J_{||}$ and $J_{\perp}$.
Numerical calculations suggest that most of the 2M Raman spectral 
density at $k = 0$ arises from the combination of triplets with 
wavevectors close to the Brillouin Zone (BZ) center \cite{Knetter}.
In that region the elementary magnons are weakly dispersive and the 
resonance situated around twice that energy reflects the 
singularities in the single particle density of states.
We estimate $r = 0.8 \pm 0.1$ and $J_{||} = 110 \pm 20$~meV, the 
error bars allowing for effects induced by the presence of ring 
exchange \cite{Matsuda2000} and finite state interaction.
A slightly anisotropic $r$ may be due to the difference in the 
lattice constants along the a and c axes as well as from the 
anisotropy of the Cu-O-Cu bonds \cite{Ohta}.
From Eq.~(\ref{Itheta}) we obtain the ratio of the coupling constants 
$A / B \approx 2.5$ in the pre-resonant regime, which would be 
compatible with an anisotropic local Cu $d$ - O $p$ excitation and 
slightly different hopping parameters \cite{Freitas2000,Muller} along and across 
the ladder.

\begin{figure}[t]
\centerline{
\epsfig{figure=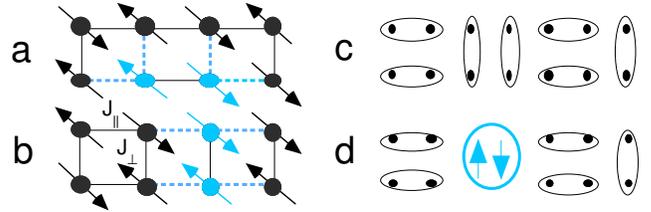,width=85mm}
}
\caption{
Cartoon showing the magnetic excitation assuming a local N\'eel (a, 
b) or a RVB (c, d) description of the ladders.
The spin exchange shown in lighter shades is along (a) and across (b) 
ladder legs.
(c) - a particular valence-bond configuration in the ladder ground 
state;
(d) - locally excited singlet state of two bound triplet excitations.
}
\label{Fig.2}
\end{figure}

\emph{Two-magnon relaxation}.--
To emphasize the 2M sharpness we compare it to the corresponding
excitation in \SrCuOCl \ which has the sharpest 2M
feature among  2D AF copper oxides \cite{Blumberg96}.
In the latter case the full width at half maximum (FWHM) is about
800~\cm-1 \cite{Blumberg96,Knoll} while
for \SrCu the FWHM is about 90 \cm-1.

For the 2D cuprates the experimental Raman spectrum shows significant
deviations from calculations within the 2D Heisenberg model.
Calculations within spin-wave theory \cite{ElliottParkinson,Canali92}
reproduce the 2M profile peaked below $3J$ which is in good agreement
with the experiments.
However, the large width of the experimental spectra has not been
reproduced within the standard spin-wave model.
The narrow calculated width of the 2M peak was found to be stable with
respect to inclusion of the higher order spin interactions
\cite{Canali92}.
Intrinsic inhomogeneities were suggested to be responsible for the
anomalous 2M width in 2D cuprates \cite{Nori}.
A recent quantum Monte Carlo calculation of Sandvik {\it et al.}
\cite{Sandvik98} reproduced a 2M profile broader
than that obtained within the spin-wave theory but still narrower than
typical
experimental spectra.
Authors of the latter work considered the broadening as support to the
earlier claim by Singh {\it et al.} \cite{Singh89} that the broadening
is due to the strong quantum fluctuations of the Heisenberg model
with $S=1/2$.
It is interesting to note, however, that for the $S=1/2$ quasi-1D
ladder structure the 2M profile is narrow.
This experimental result questions the broadening argument for 2D 
cuprates due to quantum fluctuations.
The major difference in the 2M width between 2D cuprates and the
spin ladders is due to the suppression of the magnetic relaxation 
channels in the gapped quasi-1D weakly coupled ladder system.

\begin{figure}[t]
\centerline{
\epsfig{figure=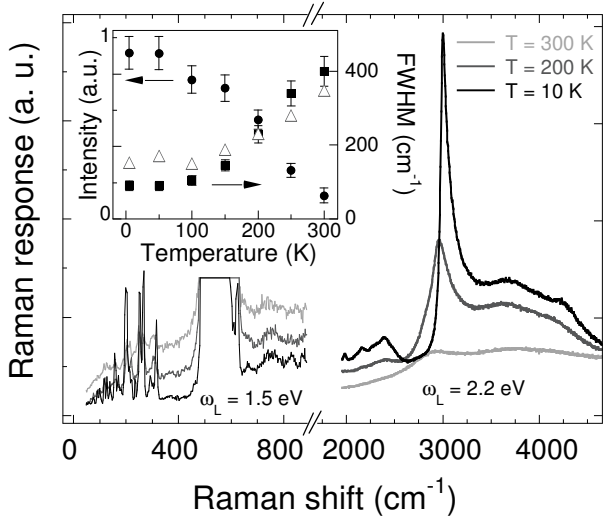,width=80mm}
}
\caption{
Temperature dependent Raman spectra in (cc) polarization for \SrCu.
Different energy scales were used for the two excitation regimes.
Left: $\omega_{in} = 1.5$~eV (the phonons around 550 \cm-1 are 
truncated).
Right: $\omega_{in} = 2.2$~eV where the 2M peak at 3000 \cm-1 is
shown for three temperatures.
Inset: the integrated intensity of the 2M peak (circles, left scale) 
and
the FWHM (filled squares, right scale).
The triangles represent the continuum intensity around 700~\cm-1.
}
\label{Fig.3}
\end{figure}

\begin{figure}[t]
\centerline{
\epsfig{figure=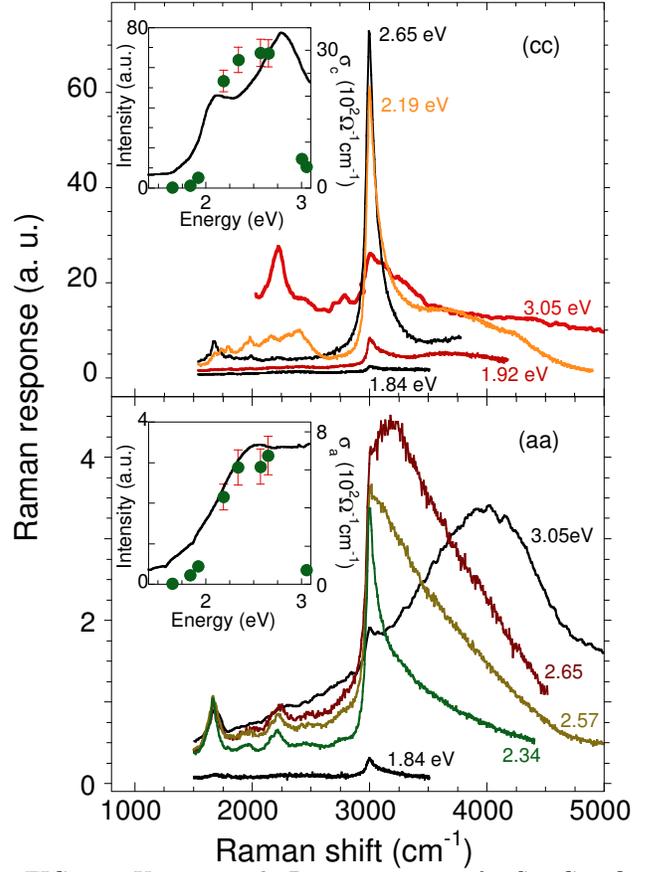,width=85mm}
}
\caption{
Upper panel: Raman response for \SrCu in (cc) polarization at T = 10 
K, for
different excitation energies.
Inset: the resonance profile of the 2M peak (filled circles, left 
scale),
and the c-axis optical conductivity (solid line, right scale).
Lower panel: The same data for the (aa) polarization; in the inset the
solid line represents the a-axis optical conductivity.
}
\label{Fig.4}
\end{figure}

\emph{Two-magnon temperature dependence}.--
The 2M peak is weak and heavily damped at room temperature (Fig.~3).
Upon cooling we notice two main features:
first, the spectral weight increases by almost an order of magnitude
and second, the 2M peak sharpens from about 400 to 90 \cm-1 FWHM at
$T = 10K$.
Because $J/k_{B}T$ remains large even at room temperature, the
observed effects are surprising.
In 2D cuprates, for example, the 2M peak remains well defined above
600~K \cite{Knoll}.

The electronic continuum shown in Fig.~3 for energies below
1000~\cm-1 gets suppressed with cooling.
The presence of low lying states at high temperatures is confirmed by
NMR \cite{Takigawa} and also in measurements of c-axis conductivity
\cite{MotoyamaOptics}.
In the latter case a gap-like feature below 1 eV develops with cooling
and the low energy spectral weight gets transferred to energies
between 2 and 3 eV.

As seen in the inset of Fig.~3 the increase of the electronic Raman
background with heating is correlated with the damping of the 2M peak.
The introduced low energy states reduce the lifetime of the magnetic
excitation by providing additional relaxational channels.
We may speculate that the states are provided \emph{via} the small 
amount of
self-doped carriers in the ladder subsystem \cite{Takahashi}.
The strengthening of the 2M seen with the 2.2 eV excitation energy
might be related to the enhancement of about 30\% of the optical
conductivity around this energy.

\emph{Two-magnon excitation profile}.--
The superexchange mechanism involves hopping of electrons to the
nearest neighbors sites \emph{via} the intermediate oxygen $2p$ 
orbitals.
Therefore we expect a change in the 2M band intensity and shape as the
incoming photon energy approaches the charge-transfer (CT) gap.
Fig. 4 shows the Raman response for \SrCu for different excitation
energies.
The 2M resonant Raman excitation profile is plotted in the insets
along with the optical conductivity data.
For both the (cc) and (aa) polarizations the resonant enhancement has 
a
maximum around 2.7 eV, about 0.7 eV higher than the CT edge seen in 
the
optical conductivity.
The intensity is small for $\omega_{in} < 2$~eV and increases
monotonically as the photon energy becomes comparable to the CT gap 
followed
by a drop for excitations about 3 eV.
The 2M profile displays an orders of magnitude increase in intensity 
as
the incident photon energy is varied in the visible spectrum.
As in the cuprate materials \cite{Chubukov} a resonant Raman coupling 
has
to be considered in order to quantitatively understand the 
experimental
observations.
Noticeably the 2M acquires an excitation dependent sideband as can be
seen in Fig. 4.
In the (aa) polarization it appears for photon energies higher than 
2.4 eV
corresponding to the edge seen in the $a$-axis conductivity
\cite{sideband}.

\emph{Summary}.--
We studied the 2M excitation at 3000~\cm-1 in two-leg ladders by Raman
scattering.
Resonant coupling across the CT gap at about 2 eV
increases the magnetic scattering intensity by orders of magnitude.
At low temperatures the 2M is a well defined resonance with a FWHM of 
$ \approx 90$~\cm-1, much sharper than the 2M band in 2D cuprates.
From the temperature dependence of the 2M width we find that the 
magnetic
relaxation is caused by excitations seen in the low-frequency
electronic Raman continuum and the optical conductivity
as the temperature is raised.
The presence of holes in the ladder subsystem is at the origin of this
continuum.
A slight anisotropy in the ladder exchange due to geometrical factors 
$r = J_{\perp} / J_{||}$ of about 0.8 is derived from the analysis of 
our data.
We show that the light coupling to the magnetic excitations is 
anisotropic and excitation dependent.

We would like to acknowledge discussions with M.V. Klein,
A.M. Sengupta, R.R. Singh and S. Trebst.

\end{document}